# Condition Integration Memory Network: An Interpretation of the Meaning of the Neuronal Design

Cheng Qian

Omnes Complexity Inc.

Cheng@omnes.xn--rhqv96g

# Abstract

Understanding the basic operational logics of the nervous system is essential to advancing neuroscientific research. However, theoretical efforts to tackle this fundamental problem are lacking, despite the abundant empirical data about the brain that has been collected in the past few decades. To address this shortcoming, this document introduces a hypothetical framework for the functional nature of primitive neural networks. It analyzes the idea that the activity of neurons and synapses can symbolically reenact the dynamic changes in the world and thus enable an adaptive system of behavior. More significantly, the network achieves this without participating in an algorithmic structure. When a neuron's activation represents some symbolic element in the environment, each of its synapses can indicate a potential change to the element and its future state. The efficacy of a synaptic connection further specifies the element's particular probability for—or contribution to—such a change. As it fires, a neuron's activation is transformed to its postsynaptic targets, resulting in a chronological shift of the represented elements. As the inherent function of summation in a neuron integrates the various presynaptic contributions, the neural network mimics the collective causal relationship of events in the observed environment.

# 1 Introduction

The nervous system plays a critical role in an animal's ability to adequately react to the environment and realize cognitive functions. The history of neuroscientific research has been a prolonged struggle to understand the nature of neuronal interactions in the brain. For instance, systematic analysis of the neuron's possible function as logical gates can be traced back to the 1940s, when the activation of a neuron was theorized to express some propositional statement (McCulloch and Pitts, 1943). The later-discovered mechanism of spike-timing dependent plasticity (STDP) further inspired the idea that the brain learns causal connections between events via synaptic changes (Roberts, 1999). Moreover, Li (2017) and Tsien (2016, 2017) have recently formulated a mathematical approach to explain biological intelligence by exploring the possible evolutionarily conserved logic in neural circuits. Nevertheless, theories of neural codes often necessitate a high level of artificial abstractions of the observed data to fit proposed algorithms that make it fathomable to an external observer. That is, in the field of neuroscience, there are often distinct interpretations of the same phenomena occurring in the brain that all lack, however, concrete theoretical and empirical support (Kaku, 2014; Brette, 2015). Without first establishing an understanding of the architectural principle of the nervous system's logic of operation, all attempts to algorithmically explain particular functions of the brain are prone to deviate from the original and primitive intent of the neuronal design. Can there be direct mapping of intuitive cognitive or motor functions to the basic property of a neural network, which would enable the formulation of an evolutionarily sound theory of this design principle? To appreciate why evolution emerged to make the neuron such a virtually ubiquitous feature in animals, it may be assumed that there must have been a universal logic that bolstered the advantage of a neuronal structure in organisms of drastically varying complexities across the spans of its evolution. In this document, the author introduces this particular novel approach to postulate this most basic logic, and then discusses the potential realization of a framework of motor-cognitive functions, in which synapses in a neural network act as distinct chronological causal connections, with neurons as the integrators and then effectors of their causal powers.

Consider, hypothetically, the initial purpose of the nervous system during evolution: the instant causation between a presynaptic detection and a postsynaptic motor actuation, realized with neurons located in various places in the organism and some form of synaptic stimulation. The activation of the detector neuron mechanically implies that of the movement as the effector is driven to act via, potentially, electrical conduction. In this case, the output of the sensory neuron is not a representational code meant to be interpreted, but a direct causal mechanism comparable to turning on a light bulb. These instantaneous connections from conditions to events, or causes to effects, are here referred to as synaptic mappings. As neurons have the capacity to integrate inputs and reflect it on their membrane voltages, the system enables an agent to react on a basis of sufficient condition. Consider a simple circuit where the reaction effected by neuron or neural population $C$ can be activated by two features in the external stimuli detected by neurons $A$ and $B$, neither of which firing alone will cause $C$ to be sufficiently stimulated and respond with an effective intensity of activation. Here, $A$ and $B$ are collectively the condition for the organism to execute the reaction, composing a biological reaction system with threshold logic. Accordingly, the presynaptic neuron brings its maximum influence to the network by maintaining a high firing rate, whereas the lack of this frequency signifies an insufficiency of the condition triggered by its own presynaptic neuron or information in the environment. Likewise, the same network can allow *A* and *D* to collectively activate another neuron, *E*, during which *C* exhibits a state of low activity through the value of its membrane potential or a low firing rate, due to *A* being the only active presynaptic condition. As will be discussed, this expression of conditional insufficiency enables the so-called continuity in the temporal structure with which the network seamlessly adapts the organism's behaviors. Effectively, the activation of *C* or *E* here depends on the specific combinations of conditions presented. Furthermore, with updates in synaptic connections, the logical relationship between the presynaptic event and the postsynaptic reaction can be wired arbitrarily. For instance, the organism can react to a small round object (*A*) combined with green leaves (*B*) with eating (*C*), or it can execute this behavior for the round object alone with increased synaptic weights from *A* to *C*. Similarly, an agent can map stimuli not to reactions but, rather, predictions on the subsequent events, applying the same logic to a causal network of declarative memories: when a bird (*A*) flips its wings (*B*), it is to fly (*C*). In this case, *C* is a prediction convergently caused by presented conditions, and the activations of *A* and *B* are transformed to *C* in the dynamic network. In the following sections, neural networks based on this operational logic will be referred to as condition integration memory networks (CIMNs).

## 2  Integration of Symbolic Conditions

In a CIMN, neurons are innately representational of memory items or motor behaviors. The neurons can represent dynamic and static graphical features such as orientations of movements, color gradient in a relative position, sounds, and other arbitrary sensory information or motor reactions, with their meaning and function effectively defined by synaptic connections. Accordingly, a group of such neurons can be seen as a collective unit (e.g. imagery of an apple) given common presynaptic conditions and postsynaptic effects. Such a unit is often referred to as a neural population or cell assembly. Furthermore, the neural network exhibits continuous dynamics, where neurons' activations are constantly transformed and shifted through synaptic connections. In other words, a story of the external world can be recalled in a form of chain-reaction effect as a train of transformations symbolically enacts its chronological evolution.

As the intensity of a neuron's firing rate determines the occasional effectiveness of its representation, episodic memories are expressed in a temporal structure of activation: Let $X_{(n)}$ be a group of excitatory neurons selected to represent some arbitrary graphical frame of events, and $X_{(n)}^{act}$ the behavior of $X_{(n)}$ in a state of sufficient firing for some unified duration. An ordered activation of all elements in a sequence $S = (X_{(1)}, X_{(2)}, ..., X_{(m)})$ passively represents a series of continuous events perceived in the environment when all $X_{(n)} \in S$ successively reaches a sufficient intensity of firing rate, denoted here collectively as $S^{act}$. $S$ is considered a memory existent in the network that the agent is able to recall independently if, with some synaptic connections and weights assigned for all $X_{(n)} \in S$, $S^{act}$ can be induced with only the initial activation of $X_{(1)}^{act}$. For the sake of simplicity, let $\phi(X)$ denote a function that synaptically transforms the level of activation of $X_{(n)}$ to that of its postsynaptic targets, and $S$ is considered a memory when

$$\left(X_{(1)}^{act}, \phi(X_{(1)}^{act}), \phi^{\circ 2}(X_{(1)}^{act}), ..., \phi^{\circ m-1}(X_{(1)}^{act})\right) \Rightarrow S^{act}$$

where the activation of the dynamic object $X_{(1)}$, representing the information of the initial time frame, causes the series to be recalled sequentially as the neural activation is transformed for $m - 1$ iterations. In other words, the sufficient activation of $X_{(n)}$ implies that of $X_{(n+1)}$, and a sequence of events emerges from the causal iterations. When this condition is satisfied, it can be said that the memory of $S$ exists and $S$ is more than a passive neural representation. As such, memory is defined by synaptic connections in the system. $\phi^{max}$ will be used to denote continued iterations of transformations, which further simplifies the above to $\phi^{max}(X_{(1)}^{act}) \Rightarrow S^{act}$.

The function of $X_{(n)}$ in the CIMN is defined by its pre- and postsynaptic groups of neurons, with its equipped synaptic weights stating its required conditions and influence, where $n$ represents a relative time frame. Despite potentially distributed in random locations, $X_{(n)}$ can be easily visualized collectively as a layer of neurons activating in such a time frame, with a form similar to that of feedforward artificial neural networks. This layer emphasizes the topology of the occasional chronological composition of a sequence of memory. Moreover, the informally defined time frame here is a small interval under which a firing rate can be derived, such that a neuron expresses a proposition as some scalable value through its intensity of firing in the dynamic system instead of fixed binary states of on and off. In view of this, such a system of representation can be understood as exhibiting a form of fuzzy logic, where a proposition can be expressed as having some degree of truth to represent uncertainty and non-absolutes based on the conditions given (Zadeh, 1962). Lastly, the representation of an event is considered emergent because a neuron is here an independent agent that does not belong to any particular memory.

The above has overviewed the continuous process of neural activation that defines the logical recalling of events as it does prediction in the hypothetical system of CIMN. The psychological nature of the notion of episodic memory is here established in reference to Tulving (2002) and Anderson (2015), in that it is recollection of past events and fundamental to narrative intelligence, and the Soar cognitive architecture (Nuxoll and Laird, 2004; Nuxoll and Laird, 2012), in that it exhibits a sequential frame-like chronology. On the other hand, the term "story" describes a more broad series of mental representation which can be imaginatively, hypothetically, or linguistically involved (Akimoto, 2018; Akimoto, 2019).

Finally, because the neuron's expression of low firing rate here signifies a lack of presynaptic conditions, the signal summations in some postsynaptic neurons can be described as consistently incomplete during some time frame. This renders certain sets of activations ineffective without a high firing rate. Whereas events in the environment often need multiple prerequisites to happen, most mapped neurons will not reach a sufficient intensity of activation due to the lack of converging evidences. With this, neurons in the defined layers can often be activated with drastically varying proportionalities. These occurrences of distinguishing activation intensities provide a basis for the following theory of information potentiality (INP), which serves as a substantial concept in this framework of biological intelligence.

# 3  Information Potentiality

### 3.1  The Continuity in the Transformations of Neural Representations

An INP is defined as an arbitrary selection of representational neurons that may together compose some frame of memory, reaction, or cognitive function, and which can be subjected to a state of incomplete activation due to the lack of stimulation or disinhibition. The INP can show partial activation, or pre-activation, which is manifested by either an insufficient number of activated neurons, or insufficient firing rates of the population such that the neurons cannot hold significant influence.

The characteristic computational function of nerve cells, signal integration under some threshold potential, has brought the capacity for logical propositions (McCulloch and Pitts, 1943; Yoder, 2009). The differences in firing rates may further expand the neuron's capacity of expression from the simple states of on and off to a measurable intensity in a dynamic system and therefore allow temporally fluctuating fuzzy logic to be utilized. In the CIMN, the variances in neurons' graded potentials and then firing rates complete a substantial design: the continuity of synaptic mappings. That is, as the proportionalities of neural activations allow all forms of potentialities to be presented in a neural network's dynamic expression, the circuits enable a smooth transition on predictions of events and the execution of tiny adjustments during movements. In a given time interval, as neurons with high firing rates directly represent events, less activated INPs express the potentialities thereof; likewise, in a single instant, potentialities can be represented by the value of the cells' membrane potential (Figure 1), though this may not be functionally meaningful. Additionally, when an element is represented by a homogeneous population of neurons, continuity can also be expressed with the number of activated neurons. In other words, a higher number of activated neuron implies a greater causal power of the current representation, functioning similarly to the variability of firing rates.

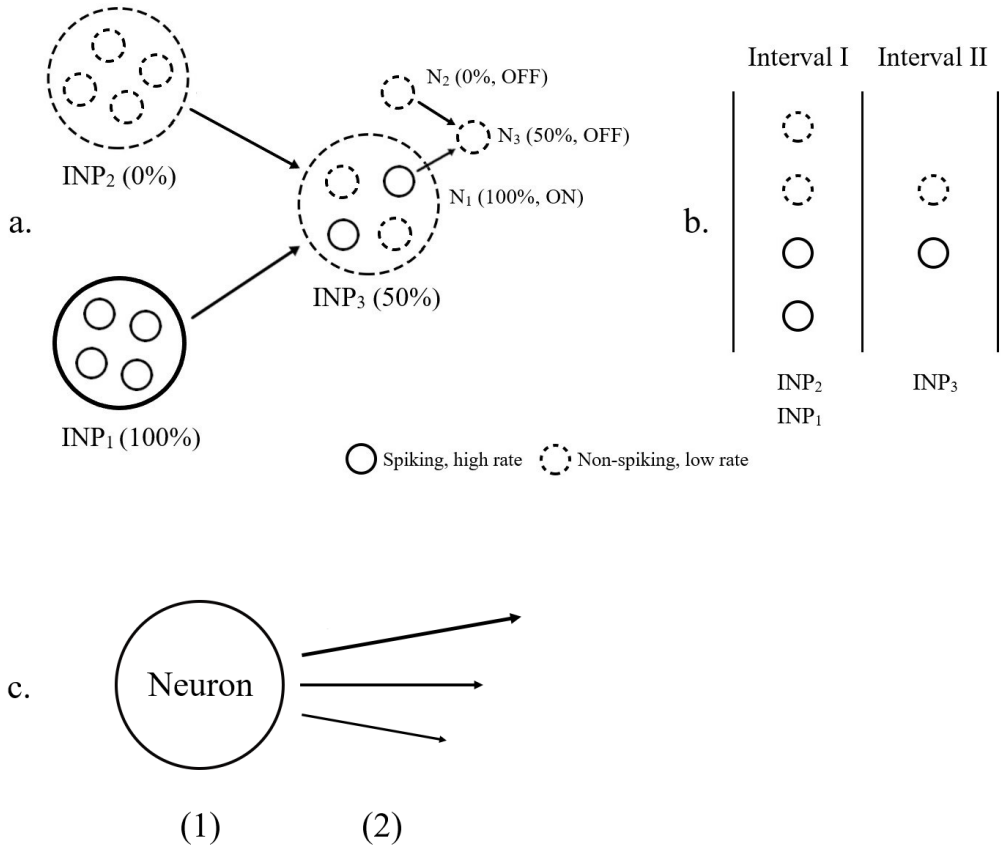

**Figure 1**: (a) A schematic illustration for the fractal-like display of the proportionalities of activation in a network. This shows the percentages of actively spiking neurons in some INPs and that of graded potentials of individual neurons *N* in some instance. Ideally, the distribution for the physical locations of neurons is inconsequential: the function of a neuron is here defined topologically by its presynaptic condition and postsynaptic influence. (b) The dynamic expression of neural networks is collectively demonstrated with the firing rates of neurons that span across time intervals. (c) Essential design of a neuron. 1. The cell's firing rate indicates the effectiveness of current representation, the sufficiency of its presynaptic condition, and at the same time its followed causal power. Accordingly, the neuron with low activity represents a less likely expectation or action given previous conditions, which naturally exhibits reduced causal influence. 2. Synaptic weights regulate and distinguish the cell's particular causal powers.

## 3.2 Implied Effects

To further explain the notion of continuity, suppose that one is attempting to take a book from the bookshelf, when suddenly the book is about to fall off; one immediately reacts and grabs the book from midair. How does this instant transition of behavior happen? The INP, or pre-activation of memory, supports an agent's rapid situational reaction. During the attempt of taking the book, the memory of the book falling off is already partially activated along with the command of catching the book, though without enough support. Then, with the sight of falling as new supportive stimulus, the INP of catching the book is sufficiently activated, causing one to act. Synaptic mappings enable this continuity of logical reactions with neurons firing in a preparatory rate due to the temporary lack of supportive conditions. Consider, also, that one looks at the sky when suddenly there is a lightning. One did not expect this to happen, but the potentiality had nonetheless been dynamically represented in the CIMN in the form insufficiently firing INPs. In this case, if an additional feature that is the dark clouds was present, the combined causal effects of the sky and the clouds would sufficiently co-activate the memory or prediction of lightning.

The changes in the presynaptic conditions yield drastically varied postsynaptic convergence and its followed activations. With this, the neural network is capable of executing an arbitrary amount of logical sequences of memory or behaviors and adapt any frame of activation to correspond with the ongoing changes in the external stimuli. Consider a train of activation defined by an arbitrary group of neurons $X_{(1)}$:

$$S = (X_{(1)}, X_{(2)}, \dots X_{(m)}) \mid \phi^{max}(X_{(1)}^{act}) \Rightarrow S^{act}$$

where $S^{act} \subseteq \phi^{max}(X_{(1)}^{act})$ is the maximum effective activation of $X_{(1)}$, effective meaning that $X_{(1)}$ can only cause neurons in this particular sequence to fire in a considerable rate. As such, the total potentialities of $X_{(1)}^{act}$ of $m$ layers can be expressed as another sequence:

$$P = (\Delta_1, \Delta_2, \ldots, \Delta_{m-1} \mid \Delta_n = \phi(X_{(n)}^{act}) - X_{(n+1)}^{act})$$

where $\Delta_n$ is the set of the less active neurons and their behaviors of insufficient firing during time frame $n$. Insufficient firing rate implies a rate below the arbitrary standard of high or sufficient activity, $X^{act}$, and $\phi$ denotes a single instance of transformation from one time frame to another which involves all postsynaptic activations of any rate. In an initiated train of activation, such potentialities lie in each layer or time frame, and yield variations when alternative conditions are presented.

# 4  On the Biological Nature of Neural Networks

**4.1  A Universal Trait of Spiking Behavior**

Across the history of neuroscientific experiments, one of the most direct characteristics observed in the nervous system, detached from any theoretical interpretation, is the response to stimuli on a basis of the intensity of firing rates. The famous publication by Hubel and Wiesel on the research of the primary visual cortex in cats provided detailed recordings on its receptive fields with the measure of maximum response, deeming some characteristics in the presented stimuli as the most "effective" for the units and regions (Hubel and Wiesel, 1962). Modern findings conclude that most sensory neurons can only respond to one highly specific type of stimulus, and the signals from these must be integrated to derive some complex properties about the environment (Frings, 2012). In a recent publication, Brette (2018) further argues that sensory neurons are in fact unfit to encode abstract information. Moreover, in terms of motor output, experiments suggest a clear linear relationship between the force of muscle contraction and the firing rate of motor neurons or the size of the motor unit (Conwit et al., 1999). The proposal of CIMN attempts to construct the potential causal mechanism which constitutionally invokes the phenomenon that the sensory system operates largely on a basis of firing intensity. By following the same school of thought, it seeks to explain the unified circuitry distribution in the animal neocortex. This uniformity has long suggested that the structure responsible for distinct higher cognitive functions in animals operates on a general design principle, which even renders brain modules capable of reorganizing cortical functions in the absence of desired early development (e.g. Mountcastle, 1957; Mountcastle, 1997; Burton, 2003; Renier et al., 2010; Klinge et al., 2010).

**4.2  Current Theories on the Brain**

The current literature on the fundamental nature of neural information processing has largely involved a somewhat unproductive debate over whether the brain uses rate coding, which implies information is encoded in the firing rate of neurons, or temporal coding, which suggests it is the timing of the spikes that convey information (Rullen and Thorpe, 2001; Ainsworth et al., 2012; Brette, 2015). Rate coding was originally inspired by the research of Edgar Adrian and Yngve Zotterman, who testified that the weight of the stimuli directly influenced the frequency of action potentials on sensory nerve fibers, while the shape of the action potential itself was stereotyped and thus unlikely to communicate (Adrian and Zotterman, 1926). It was, therefore, a common methodology to describe the behavior of sensory and cortical neurons on a basis of firing rate, which is further encouraged by its simplicity to measure experimentally. Nonetheless, the precise mechanical meaning of such behavior in the brain is, unsurprisingly, still unknown. Temporal coding also has a degree of functional resemblance in the nervous system such as the context-dependent synchronization of neuronal discharges observed in the brain (Uhlhaas, 2009), which can act as a partial argument for the importance of spike timing. It is thought that the spike timings during some window actually encode distinct symbols comparable to the binary codes in digital computers (Theunissen and Miller, 1995). Unfortunately, there is no concrete evidence or functional blueprints on how either of these two coding schemes can constitute the activity of even the simplest neuronal systems, such as the dragonfly. In the dragonfly's nervous system, the outputs of target-selective descending neurons with different directional preferences were shown to be mechanically summed at the thoracic ganglia as a form of population vector (Gonzalez-Bellido et al., 2012). In other words, these prey-detecting neurons are thought to drive motor actuation on their respectively connected wings directly, signifying an unmediated stimulus-behavior coupling (van Hemmen and Schwartz, 2008; Gonzalez-Bellido et al., 2012; Nordstrom, 2012). This leaves little space for neurons to act as an interpreter of the supposedly abstracted information that the property of action potentials encodes. The innate decoding, storage, and manipulation of data which are indispensable in theories of neural coding appear absent in this iconically efficient, sustainable system. Similarly, it is well known that cells at the sinoatrial node in the heart send action potentials on which the heart rate is mechanically dependent (Boyett, 2000; Monfredi et al., 2010). To describe this process of electrical conduction in the context of neural coding is equivalent to stating that the ignition lock cylinder in a vehicle encodes information telling the vehicle to start, which will be decoded by the internal combustion engine. This, clearly, is a conceptually inadequate and misleading way to interpret a mechanical system. Superficially, it is reasonable to assume that the firing rate represents information based on numerous experiments that show the frequency of spikes increases along with stimulus intensity. However, this assumption fails to provide a clear explanation of why this is a meaningful or efficient design or how simple, primitive organisms could utilize it. Without understanding this basis, postulating the behaviors of complex and specialized neural interactions within the brain involves much imagination and is likely to be radically inaccurate. After all, the simple increase in sensory neurons' reactions to more intense or "fitting" stimuli does not imply that information can be abstractly encoded in the brain so that one certain frequency of a neuron's spikes conveys a square shape while another frequency conveys the number "five." Similarly, that some part of the nervous system requires synchrony and precisely timed events does not imply that differently timed spikes can collectively act as a representational code.

In a perspective of evolution, the central mechanism of the nervous system should have the capacity to be simplified to a most basic and primitive form which could be utilized by early organisms that first adopted its prototype design. And, ideally, the first nervous system was of an intuitive logic, with which neurons and synapses are able to perform motor-cognitive operations as their natural function without relying on abstract algorithms. The author presumes that the gradual process of evolution should not allow the latter to have existed in parallel to the birth of the nervous system. That is, complex neuronal algorithms are more likely to be developed gradually with respect to the most primitive functional logics of neurons, which remain undiscovered. Such algorithms, nonetheless, are a prerequisite in any theory of rate and temporal coding by definition. Without highly abstract neural algorithms, no information can undergo the "decoding" process which is required for any system that relies on the metaphor of neural coding, as it is therein the only means by which the nervous system can act on any stimulation (Somjen, 1972; Jazayeri and Movshon, 2006; Byrne et al., 2014; Brette, 2018). Indeed, a system that can only work favorably for an organism once it has reached a dramatic level of complexity and organization is not a system that is sound from an evolutionary perspective. Numerous theories of neural coding, likewise, are not sufficiently adequate to describe the brain's function and suffer overall from the lack of empirical support (Perkel and Bullock, 1968; Brette, 2015; Brette, 2018).

State-of-the-art research on the visual processing of the human brain now allows a high resolution bidirectional translation between one's neural activation pattern and the perceived imagery (e.g. Roelfsema et al., 2018; Shen et al., 2019; Ren et al., 2021). Such practicality is achieved with observation on the distinct intensities of neural activation, which comparatively suggests a form of direct symbolic representation by neurons. This brings us to another set of general coding schemes, population coding and sparse coding—both of which suggest that information can be represented jointly by the activation states of neurons. Theories such as the Hierarchical Temporal Memory (HTM) have gained popularity in recent years, which states that the sparse activations of cortical neurons represent spatial and temporal patterns (Hawkins et al., 2019). This is supported by machine learning simulations and the increasingly evidential phenomena suggesting that information is represented with sparse firing distributions in the cerebral cortex (Barth and Poulet, 2012). However, practically, population and sparse coding often lack the critical capacity to be universally appliable to cognitive and motor procedures at a general level. That is, a basic logic of neural transformation that can allow it to approach every cognitive and motor function, however complex or simple, still seems to be unattainable.

Eventually, the theory of CIMN attempts to complete the design of neuronal representation by population, under which the phenomena of rate coding exist as a means to achieve the desired representational transformations. An argument here is that the sole purpose a neuronal representation should be to physically invoke its followed activations, instead of showing some external observer what the mind is visualizing at the moment. The representation is caused by synaptic stimulations, and its own influence and effect must rely on what it can further stimulate through synapses as they are the primary causal mechanism in the brain. In other words, it should have the capacity to synaptically drive and ultimately create another set of representation. If this process is continuously iterated, as described with the CIMN, declarative memory can be realized as a chain of logical inference, or a movie with successive frames. The interpretable synaptic transformation in the CIMN is hypothesized to be able to simulate any cognitive and motor procedures that rely on the basic logic of cause and effect. Furthermore, the behavior of condition integration is a reasonable choice of the causal mechanisms for neuronal representation because of its direct biological mapping. Synaptic connections determine what each representation is causal of, the synaptic weight dictates the specific amount of causal power, and the membrane voltage of a target cell shows the sufficiency of the joint causal powers along with its followed exhibition of grand excitatory postsynaptic potentials. The role of inhibitory neurons can also be easily inferred in such a setting: the chronological mapping between events is further polished with the implementation of inhibitory neurons to cope with the drastic non-linearities in the world, where the presence of some features compromises the possibility of otherwise likely events to subsequently occur. Such computational logic and method of expression are intrinsic to the neuron and necessitate no artificially defined algorithms to function. With this, the theory seeks to finally explain the functional nature of the neuron in a manner which is also plausible in the perspective of evolution.

# 5 Learning and Initial Computation

## 5.1 Potential Mechanism of Learning

A core hypothetical concept here is that synaptic connections can be assigned to a pool of neurons in a manner that, with a proper selection of initially activated neurons, the followed evolution of the network always symbolizes a logical story. This is the defining characteristic of the CIMN that enables the logical variabilities of reactions and predictions. To successfully achieve this, the agent must first be able to learn and extract temporal patterns in the external world. Furthermore, the pattern connections must be reinforced in the face of repeated encounters through learning mechanisms such as the STDP, or what one may refer to as a Hebbian-fashioned learning. In this section, a low-level computational simulation on some of the basic functions of such neural networks will be briefly demonstrated with a newly developed library CoreNEURON (Kumbhar et al., 2019).

Unsupervised learning is necessarily involved in the CIMN, where events are modeled without biases. To learn in a presented environment, every element or feature in an initial state of the world, represented by a neuron, establishes an excitatory connection to every element in the subsequent state. Therein, the weight of each such connection is increased every time the particular sequence is repeated. This would ultimately result in every neuron, equipped with synaptic connections of various strengths, becoming a condition that pushes on the next state of the world as it activates and causes varying influences to the postsynaptic targets. The simple system can then take in an arbitrary state of the world and synaptically map it to the most likely outcome before such is actually perceived, that is the postsynaptic neurons with the highest spike activities due to the convergence from presynaptic conditions. With this, every synaptic weight can be understood as representing the level of potentiality for some particular change given the presence of an event, which is further integrated with other concurrent spike trains in a manner comparable to elements in the world bringing collective causal effects. In other words, the amount of repetitions during learning dictates the network's interpretation of how likely the presynaptic neuron contributes to one of its postsynaptic expectations. A neuron for an event and a synapse for its potential future—the neural network functions without algorithmic rules to directly mimic the causal mechanisms of the world.

Consider a detection system that receives input in identical time steps and activates either neuron $i$ for stimulus $A$ or neuron $j$ for stimulus $B$ in each step, where the spike trains of $i$ and $j$ are uniform in both timing and duration when the detection system is their only source of input. Two successive time steps with sequence $AB$ or $BA$ are considered a discrete iteration with which synaptic weights can be updated. Neuron $i$ has been initially established as a presynaptic neuron of $j$, and the change of its weight $W_{ij}$ after $n$ iterations of $AB$ and $m$ of $BA$ will be defined by $\Delta W_{ij} = nr - mr$, where $r$ is an arbitrary learning rate in reference to the current weight. In other words, the stimuli presented in a desired sequence will increase the connectivity of the neuronal representation of this chronological pattern, which will be decreased if the stimuli come in a reversed order. The following demonstrates computationally that the classical mechanism of STDP, which functions on temporal correlations of spikes, may naturally approach such a learning rule.

### 5.2.1 Sequence Learning at Neuronal level

An artificial Integrate-and-Fire neuron, developed after van Elburg and van Ooyen (2009), is used for the representational neuron in the simulation for the sake of generality. A modified STPD learning algorithm derived from Chadderdon et al. (2012) and Davison and Frégnac (2006), as part of the development of NetPyNE (Dura-Bernal et al., 2019), is utilized. The detection system sends 10 uniform spikes in an interval of $10ms$ through non-plastic connections for each stimulus in a $100ms$ time step. The neurons $i$ and $j$ are given a $20ms$ learning window with a maximum weight adjustment of 0.05 per inter-spike interval, and the presynaptic signals from $i$ travel to $j$ in $\sim 30ms$. The simulation shows that the efficacy stably increases when the neurons are stimulated by the desired pattern, and largely vice versa (Figure 2, 3). Additionally, similar learning curves are resulted with adjustments of parameters, whereas the spike count from the detection system is gradually reduced from 10 to 1 and lessened travel times are implemented. That is, the form of these learning curves is considerably preserved in alternative parameter settings given $W_{ij} < \sim 0.8$ for the reversed sequence. However, the efficacy would generally increase regardless of the sequence when the weights are set above this threshold, soon before which the neuron begins to independently cause the postsynaptic target to fire in rate higher than itself—effectively outweighing the influence from the detection system with gain. This may further imply a generalized STDP learning structure which is not reliant on precise parameter settings.

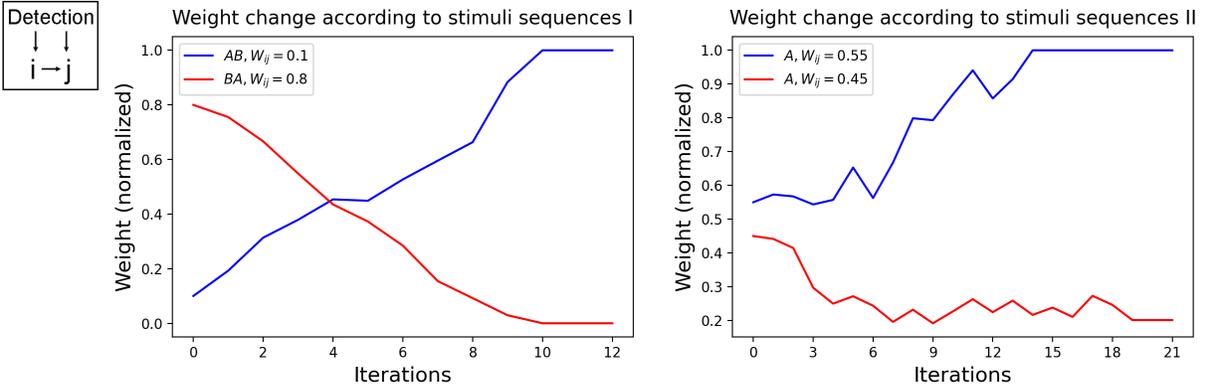

**Figure 2**: (I) With an initial weight of 0.1, the repeated iteration of the desired sequence *AB* yields a continuous increase in efficacy which stabilizes at the maximum weight, 1. Conversely, a reversed sequence *BA* brings a stable decrease. (II) Until $W_{ij}$ reaches a certain threshold, repeated encounters with the absence of the postsynaptic event still decreases the expectation. When the weight becomes of sufficient value, however, recalling the event causes the connection to be reinforced instead.

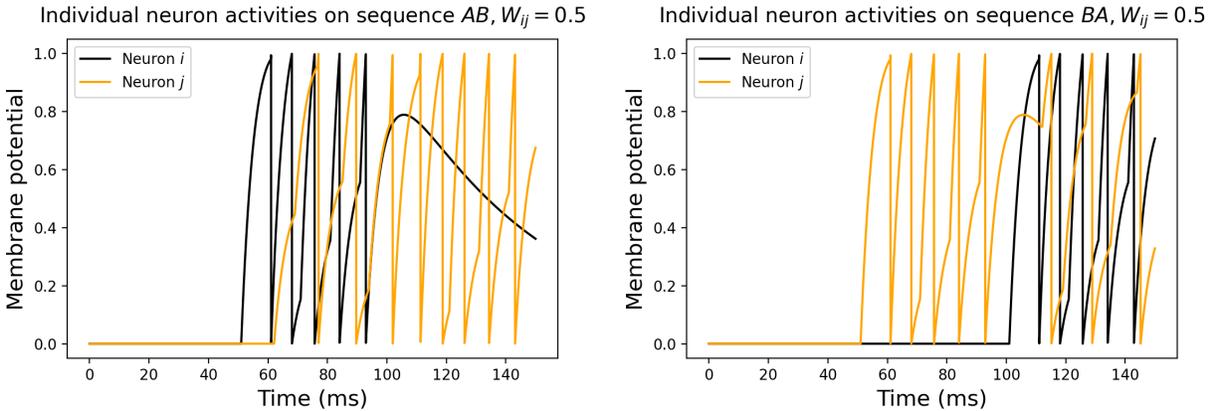

**Figure 3**: Analytic values of *i* and *j*'s membrane states in a parameter setting of 5 spikes from detection in $50ms$ time steps and uniform $1ms$ travel time. These yield separately a final efficacy change of $+\sim 0.09$ and $-\sim 0.03$ through the implementation of STDP. STDP is a temporally asymmetric process that alters synaptic strength based on the arrival time of a presynaptic spike and the cell's activation. The synapse is strengthened if a presynaptic spike arrives shortly before the postsynaptic neuron activates and weakened when it arrives shortly after. For a more detailed overview of STDP in modern research, see Sjöström and Gerstner (2010).

The above is an example of how stimuli-guided learning may be realized with temporally asymmetric Hebbian plasticity in a closed system of CIMN. In this case, the causal connection is weakened through long-term depression (LTD) as the agent observes the events occur in a reversed order, which implies these are potentially not causal. In other words, if the weight assumes that an apple is to appear after some cue, witnessing the apple before the cue lowers future expectation. See Roberts (1999) for a detailed analysis on the biophysical plausibility of a similar learning rule.

### 5.2.2 Expression of Continuous Predictions

The parameterized system will now learn the evolutions of a few simple dummy events. 10 artificial neurons indexed 1~10 will each represent a unique element or memory item in the dummy events, which are given by $M \times N$ matrices

$$E_l = \begin{bmatrix} A_{11} & \cdots & A_{1N} \\ \vdots & \ddots & \vdots \\ A_{M1} & \cdots & A_{MN} \end{bmatrix}$$

where $N$ is the number of time frames and $M$ the element count in a frame, with 0 as an empty entry. Events exhibit intrinsic causality which is meant to be assimilated by the learning mechanism. With a detection system

$$D: E_l \rightarrow S_l$$

the CIMN yields its passive representation $S_l$ for the $l$th event in real time. Then, through repetitions of such input, weights are assigned to the network such that it predicts the followed evolution of the event given only input of the initial time frame (Figure 4, 5). In this case, the time it takes for the subsequent elements in each time frame to successively reach their peak firing rates is significantly shorter than that in the real time, hence the function of prediction. If the mapping time is not instant and takes as much time as real events, the predictive purpose must be less effective. Additionally, though the learning rates are not identical throughout the chronological compartments of the neuronal representations, a general and continuous efficacy increase through long-term potentiation (LTP) towards the direction of desired event recall is maintained.

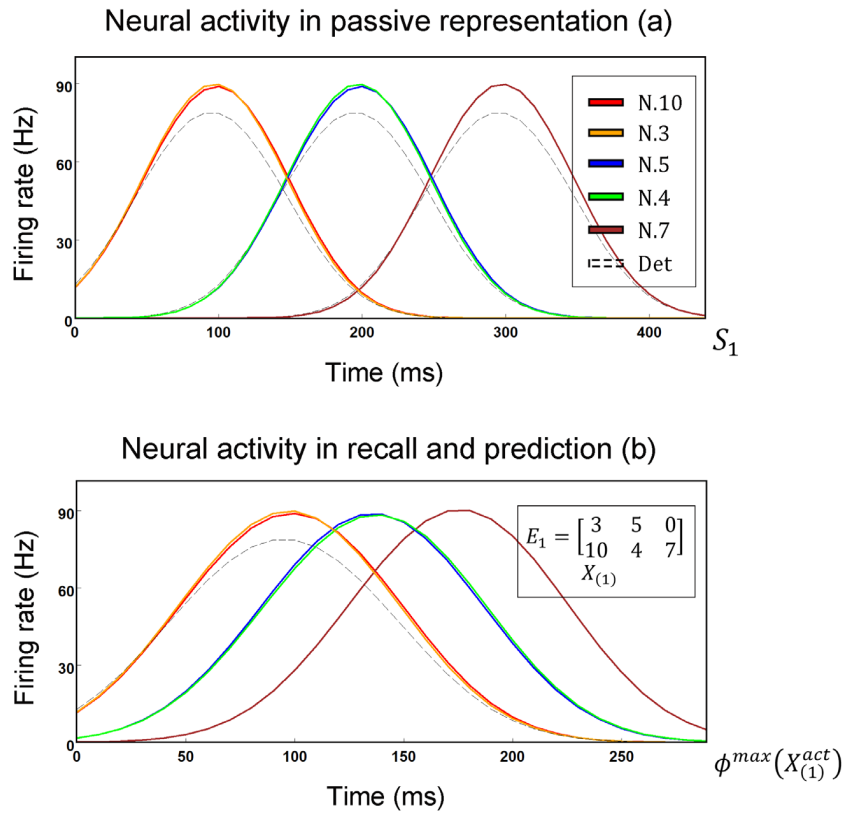

**Figure 4**: The patterns of activation with firing rates in Gaussian distribution. Activity of the detection system is shown by dotted line. (a) Representation in real time: Synchronized activation pattern from all neurons due to uniform input in each time step, without weights assigned. (b) Recalled events with an initial cue: After learning, the CIMN predicts the followed evolution of the event with an initial frame input. The amount of potentiality as measured by firing rate gradually increases for each element and peaks at the most fitting time frame (in reference to the activation of other neurons), which is also the maximum rate allowed in this setting.

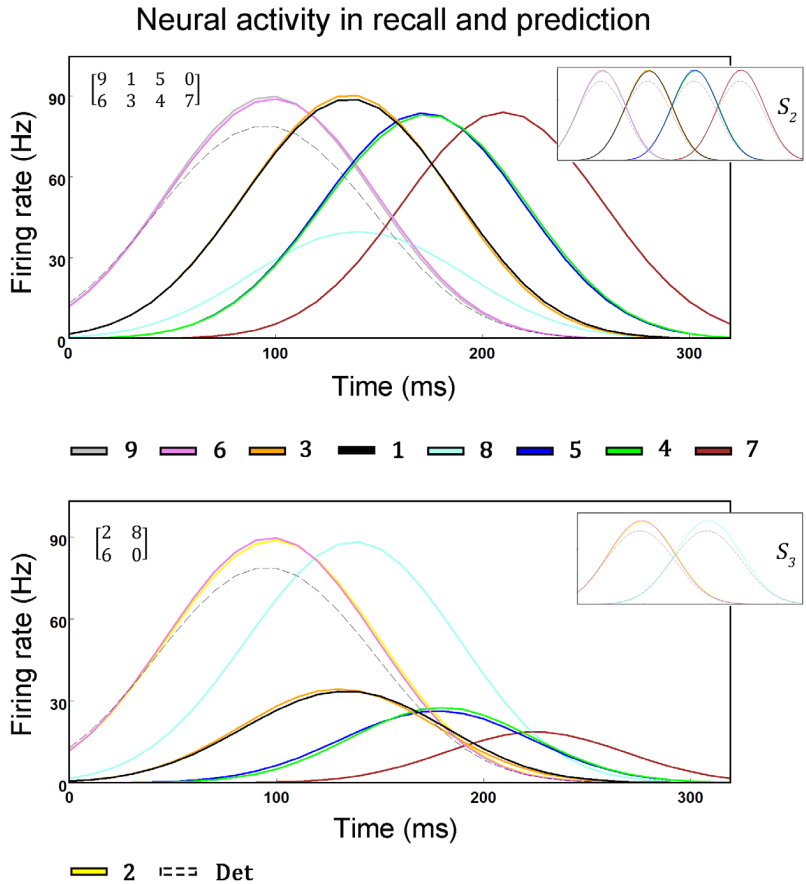

**Figure 5**: The system reacts with the set of weights from learning as an interpretation of the causal influence of each input stimulus. A CIMN dynamically outputs not only the most likely events but also the potentialities with lower probabilities of realization due to the lack of predicating conditions. Whereas two conditions, neurons 9 and 6, are sufficient to drive the subsequent scenes, 6 alone results in a less effective representation thereof as shown in the second graph. Similarly, 8 is a potentiality in the presence of 6 and yields a higher probability with the additional support of 2. Do the unactualized potentials actually manifest themselves in the physical world? This is rather a philosophical question.

This section has discussed the CIMN's representational dynamic at the neuronal level with a micro functional display of the network's logic. A temporal probability distribution of all likely events is generated in a causally informed system of neural network, presenting their most likely order of arrival according to the initial condition. Events with high chronological proximity are considered as a collective frame, or layer, composed by neurons as independent agents. Suitably, the content of the story brought by a logical train of activation with multiple time frames can be drastically varied when the composition of the initial frame is changed. Nevertheless, in real applications, more elements should be incorporated with complex data that reflects high resolution events, from which a realistic measurement of performance can be derived.

# 6 Behaviors and Movements

## 6.1 Mapping Stimuli with Reactions

The neurons' causal nature is here aligned with critical cognitive functions as they possess symbolic meaning. The neurons are events and at the same time conditions to their potential states, as inevitably are elements in the real world. Accordingly, synapses are objects that act on time, representing these potentials in an environment where things have various tendencies to change. The same logic must also be applied to movements, as the sensory-motor transformation is an evolutionary priority of the nervous system. In a CIMN, actions share the same functional principles with declarative memories and predictions, and these exist virtually interchangeably.

As aforementioned, a set of sensory inputs can be mapped to actions through the neural network where synaptic connections and their respective weights dictate how the inputs should convergently activate some actions while also keeping certain neurons firing in a preparatory rate. One can imagine waiting in front of a red traffic light, where one begins to prepare to drive when the light becomes amber and executes the movement when it is green. During this wait, the neurons that eventually push the action fire in a preparatory intensity, gradually increasing in the three stages of traffic light colors as the adequacy of the evidences change. The potentiality of driving gradually increases and peaks when presented conditions bring the highest support, as is a natural product under the logic of CIMN. In this case, the traffic device (without colors) and the road are the common features which prepare and partially activate the related actions. A similar concept is the well-known "readiness potential" popularized by the experiments of Libet et al. (1983), where it was argued that the rise in neural activity before an individual's conscious awareness served as an initiation or preparation for the followed voluntary movement (recent studies by Alexander et al., 2016 and Schultze-Kraft et al., 2016 have nonetheless suggested that such behavior is not unique to motor processes). Single-neuron recordings further described the evolution of the gradually increasing firing rate prior to the conscious execution of the action (Fried et al., 2011). The recorded progressive increase in the firing rate that peaks at the action, shortly before which the agent notices such intention, is consistent with the ideological nature of the CIMN. Whether or not modulated by other factors, firing intensity is here an indicator of condition sufficiency and a direct neural signature of causal power.

## 6.2 Close-loop Control

For the organism's actions to be sufficiently adequate and flexible, the behaviors being executed must also be constantly adjusted with respect to newly received stimuli and sensory feedback to construct a closed-loop control. Therein, a high number of neurons can provide seamless mapping from every tiny change in stimulus to the corresponding adjustments of reaction and the pre-activations thereof. A train of actions is varied at any layer with the change in external stimuli or sensory feedback to precisely fit a situation, whereas every potentiality immediately comes into effect once the conditions become sufficient. Consider a sequence of feedback control where the set of external stimuli $\lambda^{ext}_{(n)}$, the concurrent state of the agent $\lambda^{int}_{(n)}$, and their immediate subsequent steps $\lambda^{ext}_{(n+1)}$ and $\lambda^{int}_{(n+1)}$, are synaptically mapped to their respective movements and the potentialities thereof:

$$S = \left( \phi \begin{pmatrix} \lambda^{ext}_{(1)} \\ \lambda^{int}_{(1)} \end{pmatrix}, \phi \begin{pmatrix} \lambda^{ext}_{(2)} \\ \lambda^{int}_{(2)} \end{pmatrix}, \dots, \phi \begin{pmatrix} \lambda^{ext}_{(m)} \\ \lambda^{int}_{(m)} \end{pmatrix} \right) \Rightarrow \left( X_{(1)}, X_{(2)}, \dots, X_{(m)} \right)$$

where $X_{(n)}$ is the effective movements executed for $\lambda^{ext}_{(n)}$ and $\lambda^{int}_{(n)}$, the flexibility of which is here given by

$$P = \left( \Delta_1, \Delta_2, \dots, \Delta_m \mid \Delta_n = \phi \begin{pmatrix} \lambda^{ext}_{(n)} \\ \lambda^{int}_{(n)} \end{pmatrix} - X_{(n)} \right)$$

stating how the agent could have moved with changes or alternations in the presynaptic condition. This concludes every possible set of reactions an agent can make according to the diverse sequences of stimuli. Figure 6 is a schematic illustration on the synaptic mappings and the evolution of the system.

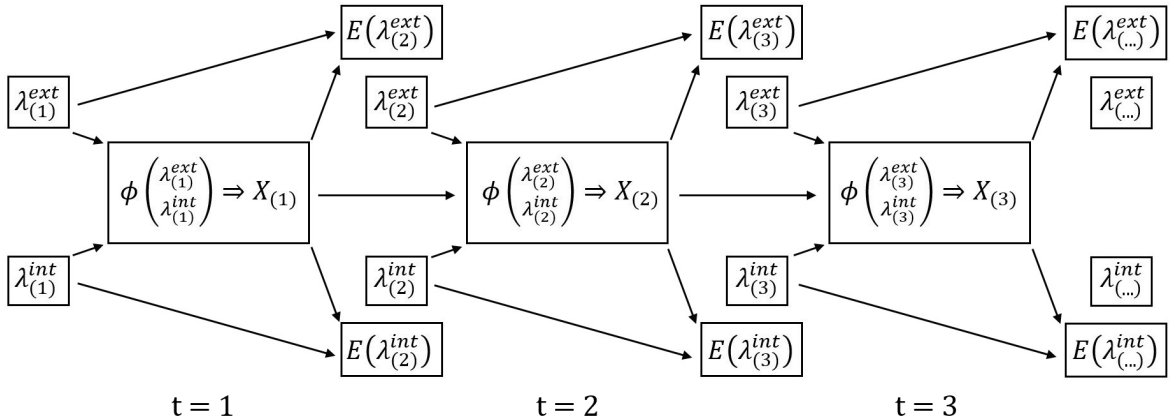

**Figure 6**: A simple closed-loop control expressed in arbitrarily timed steps, $t = 1, 2, 3$. $\lambda_{(n)}^{ext}$ is a collection of input stimuli including sensory feedback, and $\lambda_{(n)}^{int}$ is the internal representation of the agent's body position. $E(\lambda)$ is the expectation on future states, with which prediction errors can be derived. Each arrow represents a synaptic mapping that exhibits causal influence, and the integrated conditions bring emergent sets of activation that cause effective movement $X_{(n)}$. For reference, one may consider the action of picking up a cup takes to $t = 20$ steps.

Finally, the generality of the logic of condition integration allows neurons with various functions to be of direct causal influence. For instance, a visual scene and the motor command of turning to the left, when activated synchronically, can be considered one emergent frame that will be synaptically transformed to the expectation about the new scene when the agent's body turns to the left. As such, the neural network integrates the causal powers from diverse sources of information which can involve both declarative memories and motor plans.

# 7 Enabling Further Research

The author has discussed a set of novel ideas on the function of the nervous system without, however, actualizing their vast computational potential. Although the basic casual mechanisms described here are highly intuitive and likely trivial to simulate, concrete modelling is nonetheless required to validate their applicability and realize systematic experimentations. Thus, the purpose of this early work is partially to initiate further development on this theoretical approach to reconstruct biological intelligence.

Certainly, this architecture of neural network design, which is aimed to be a platform to harbor a vast selection of modular cognitive applications, would take great effort to develop and test. However, if the logics used here do follow the biological nature of the nervous system, the architecture could become a highly effective framework for the creation of an artificial intelligence. The implications of this can be seen much before technology allows detailed analysis on the working of the brain at a lower level.

# 8 Summary and Discussion

## 8.1 Declarative Memories through Synaptic Transformation

Several experimental concepts on the primitive function of neural networks are expressed in this document, which hypothesized the principles for the stimulation-reaction biological intelligence. It discusses the potential to create and study a functional framework, the condition integration memory network, which models continuous motor-cognitive processes. In the CIMN, selecting or composing a frame of event generates its followed chronological evolution. Changing the composition of this frame with alternative presynaptic conditions results in a shift of the convergences in the postsynaptic expectations. The capacity of the network is defined by the possibilities of such variances on the train of neural activation.

In terms of symbolically reenacting patterns observed in the environment, the author theorizes that the neural network stores causal relationships with its diverse connections. The neuron represents a symbolic element or memory item; when different neurons activate simultaneously, a collective frame of event is on display. Synaptic connections, furthermore, state the potential effects of each element. The network calculates the collective causality of perceived elements in the current scene through the integration of conditions, namely the neuron's innate summation process. A neuron now simultaneously represents part of an event and acts a condition as it is equipped with synaptic connections, the efficacies of which define the amount of its supposed causal contributions to the numerous subsequent events. Accordingly, a neuronal representation is considered effective with a high firing rate, which indicates the sufficiency of its presynaptic conditions and at the same time ensures its postsynaptic contributions. Propositional insufficiencies and uncertainties, likewise, are dynamically expressed in a network's behavior in the form of low firing rates. Nevertheless, the firing rate here is not a representational code to be read by an interpreter or observer, but rather a direct mechanical response. Consider a scenario where a coffee cup and the hand on the coffee cup together co-activate the memory of drinking coffee, but neither one of the two elements would trigger this when presented alone. In the latter case, drinking coffee is an unlikely expectation expressed in the network, though mechanically ineffective to the agent due to the lack of its postsynaptic influence. Furthermore, a temporal structure of activation can be created with which to exercise continued recall: as the postsynaptic neural populations converged onto, now a new frame of events, become the conditions for their own postsynaptic targets, the iteration continues and forms a chronological chain of activation. This allows events to be recollected in an orderly or causally informed fashion. Lastly, an element represented by a neuron can serve as evidence for any postsynaptic expectation. For instance, seeing the round shape may, to an inconsiderable extent, activate the expectations of encountering an apple and, at the same time, that of a baseball. The unbiased use of features as evidences ensures the capacity of generalization, which may give rise to cognitive phenomena such as the perceptual set, where the same element functions differently in varying contexts (e.g. Allport, 1955; Bruner and Minturn, 1955; see McLeod, 2010 for a conclusive review).

## 8.2 Hebbian Learning

The concepts surrounding Hebbian learning often assume that synaptic connections are a means to establish correlative properties of information, such as auto-associations (Hebb, 1949; Allport, 1985). The author argues, however, that the connections act not on the dimensions of plain associations, but the dimension of time. It is suggested that the neural system is causally informed and naturally extracts chronological patterns in the observed environment, instead of simply clustering related subjects.

## 8.3 Achieving Efficient Reactions

With regard to the evolutionary background, the author has argued that hard-wired connections in primitive organisms would bring condition-based intelligence so that a stimulus directly implies a reaction, defined by monosynaptic connections. These reflexes are of a highly efficient nature. As the mechanism of graded potentials allows more than one condition to be integrated and thereby acts as a buffer for potential reactions, this modular system can be further partitioned virtually indefinitely into highly specialized sensory neurons responsible for distinct stimuli and small motor units for sophisticated movements. Naturally, a low firing rate renders that the motor does not effectively respond with an insufficient sum of conditions. Likewise, it can be assumed that pre-activation should allow the quickest instinctive reactions to stimuli possible under the biophysical capacity of the neuron, whereas the reaction speed directly correlates with the speed of the chemical or electrical synaptic stimulation. An animal may be able to stare at a subject to maintain the activation of conditional neurons and immediately react given another condition, avoiding any inefficient abstractions of information processing by reason of the continuity exhibited in the dynamic network.

**8.4 Special Circuits**

Certainly, numerous cognitive abilities may exist outside the scope of the basic behaviors of network and necessitate special circuitry design. For example, recent experiments show that the ability of recognizing orders of events can be removed with damage to the hippocampus or the medial prefrontal cortex (DeVito and Eichenbaum, 2011), suggesting it to be a modular function. How can, therefore, a neuron be only activated when stimuli are received in a specific order? Likewise, in what ways can a cognitive system achieve reality monitoring, the capability to differentiate real and imagined events which varies drastically among individuals (Buda et al., 2011; Nahari, 2018)? Moreover, the past 25 years have seen dramatic advances in the research of visual attention, which is reliant on highly established neurological circuits (Carrasco, 2011). Aside from the implementation of attention, the mechanisms of language, a marvel of human evolution, can only be realized with specialized neural systems (e.g. Delogu et al., 2019; Zlatev, 2008; Beeman and Chiarello, 2013). As an attempt to construct a neuronal framework that harbors distinct cognitive functions and potentially replicate human-level reasoning, the capability of the logic of condition integration must be put to further test.

In conclusion, the neural network is hypothesized to be a highly logical and efficient system that relies little on the commonly supposed processes of neural encoding, decoding, and manipulating of received data. The integration of diverse causal powers may be the natural function of the nervous system, hence the numerous synapses of a neuron. Synaptic weights, likewise, are the system's interpretation of the amount of causal influence of each represented element. With this, learning allows the adaptive synapses to model the chronological evolution of the world. Accordingly, logical thinking and behavior may also be manifested with a loop of such integrations. Aside from the attempt to inspire new ways to investigate the nature of the neuronal constructs, this document is aimed to introduce a sound framework for artificial intelligence. Can a pool of neurons be assigned connections such that, after initially activating a proper selection of neurons, a logical story is always generated? Certainly, further research and simulations are required.

Roberts, P. D. (1999). Computational Consequences of Temporally Asymmetric Learning Rules: I. Differential Hebbian Learning. *Journal of Computational Neuroscience, 7*(3), 235-246. doi:10.1023/a:1008910918445

Roelfsema, P. R., Denys, D., and Klink, P. C. (2018). Mind Reading and Writing: The Future of Neurotechnology. *Trends in Cognitive Sciences, 22*(7), 598–610. doi:10.1016/j.tics.2018.04.001

Rullen, R. V., and Thorpe, S. J. (2001). Rate Coding Versus Temporal Order Coding: What the Retinal Ganglion Cells Tell the Visual Cortex. *Neural Computation, 13*(6), 1255–1283. doi:10.1162/08997660152002852

Schultze-Kraft M, Birman D, Rusconi M, Allefeld C, Görgen K, Dähne S, Blankertz B, Haynes JD (2016) The point of no return in vetoing self-initiated movements. *Proc Natl Acad Sci U S A 113*:1080–1085. doi:10.1073/pnas.1513569112 pmid:26668390

Shen, G., Dwivedi, K., Majima, K., Horikawa, T., and Kamitani, Y. (2019). End-to-End Deep Image Reconstruction From Human Brain Activity. *Frontiers in Computational Neuroscience, 13*. doi:10.3389/fncom.2019.00021

Sjöström, J., and Gerstner, W. (2010). Spike-timing dependent plasticity. *Scholarpedia, 5*(2), 1362. doi:10.4249/scholarpedia.1362

Somjen, G. (2012). *Sensory Coding in the Mammalian Nervous System*. Springer Publishing.

Theunissen, F., and Miller, J. P. (1995). Temporal encoding in nervous systems: A rigorous definition. *Journal of Computational Neuroscience, 2*(2), 149–162. doi:10.1007/bf00961885

Tulving, E. (2002). Episodic memory: From mind to brain. *Annual review of psychology, 53*(1), 1-25. doi:10.1146/annurev.psych.53.100901.135114
39